\documentclass{elsart1p}

\usepackage{epsfig}

\usepackage{amssymb}

\begin{document}

\begin{frontmatter}



\title{Collective flow 
at intermediate energies}


\author{J.~{\L}ukasik and W.~Trautmann for the INDRA and ALADiN Collaborations}

\address{Gesellschaft f\"ur Schwerionenforschung mbH,\\
D-64291 Darmstadt, Germany}

\begin{abstract}
We present results of a flow analysis for the set of reactions of $^{124,129}$Xe 
projectiles and $^{112,124}$Sn targets at incident energies 100 and 150 A MeV
studied with the INDRA detector at GSI. The dependence on centrality and on $p_t$ 
of the directed and elliptic flow are determined for isotopically selected reaction 
products with $Z \le 3$. The flow parameters $v_1$ and $v_2$, in general, follow 
expected trends but isotopic effects are small.
\end{abstract}

\end{frontmatter}

The high accuracy that can be achieved in the experimental determination of 
flow parameters from data collected with advanced 4$\pi$ detection systems has
recently been demonstrated in a combined analysis of INDRA and FOPI data 
\cite{and06}. Excitation functions for directed and elliptic flow were constructed
for $^{197}$Au + $^{197}$Au reactions over a large range of incident energies 
extending from 40 A MeV to several A GeV using the data measured at SIS by the
two collaborations~\cite{lukasik05,andronic05} but also data taken at the 
AGS~\cite{pinkenburg99,bmunzinger98}. 

In this analysis, corrections were applied for the reaction plane dispersion 
using the random sub-event method, assuming a non-isotropic 
Gaussian distribution of sub-Q-vectors and taking into account the correlation 
between sub-events~\cite{and06,luk_iwm}, but also for other 
experimental effects such as 
losses due to the finite detector acceptance and multi-hit events. The agreement 
reached for the INDRA and FOPI data in the overlap region 90 to 150
A MeV covered by the two data sets is within 5\% to at most 10\%.

We believe that data with this level of precision can be useful for the 
investigation of presumably small isotopic effects related to the symmetry term 
in the nuclear equation of state, with the aim to reduce its present uncertainty 
at densities beyond the saturation density~\cite{ditoro06,chen07}. For example,
the recent analyses of the isospin diffusion data measured at Michigan State 
University \cite{tsang04} have shown that the result deduced for the 
stiffness of the 
symmetry energy is very sensitive to the chosen interaction and to the type of
nucleon-nucleon cross sections adopted for the transport model 
description~\cite{chen07,tsang04}. Flow observables are generally sensitive
to these ingredients and thus provide opportunities for testing the overall
consistency of the model description. Additional sensitivity to the 
isospin-dependent part of the interaction can be expected from data for reaction
systems with varying isotopic compositions as well as for analyses for pairs
of reaction products with opposite isospin (mirror nuclei).

\begin{figure}		

    \centering
    \epsfysize=8.0cm

   \epsffile{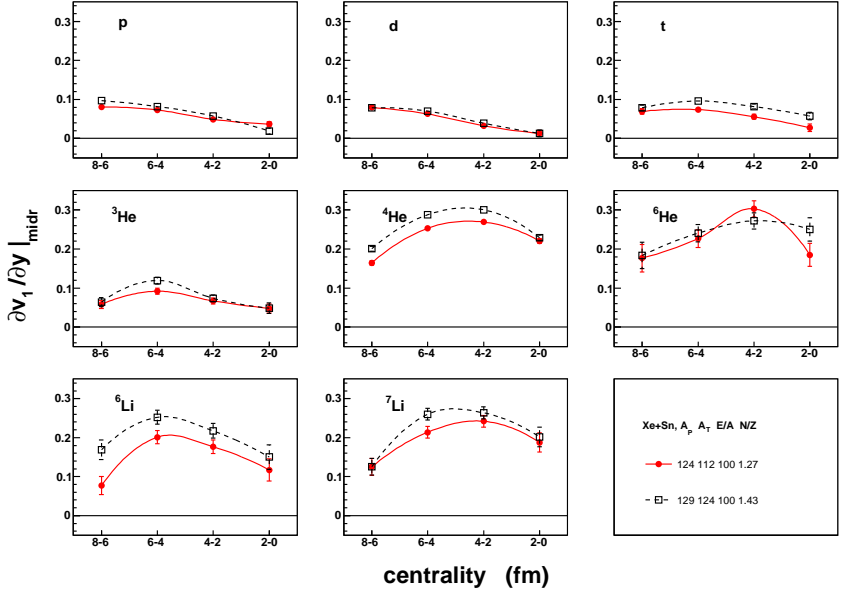}

\caption{Slopes of directed flow $\partial v_{1}/\partial y$ at mid-rapidity
for eight reaction
products with $Z \le 3$ integrated over $p_t/m > 0.1$ as a function of centrality
for the most neutron rich (open squares, black) and the most neutron poor
(dots, red) reaction systems at 100 A MeV.
}

\label{fig:v1}

\end{figure}

The data used for the present analysis were obtained by the INDRA and ALADIN 
collaborations in measurements using $^{124,129}$Xe projectiles of 100 and 150 
MeV per nucleon delivered by the heavy-ion synchrotron SIS at 
GSI. Isotopically enriched targets of $^{112,124}$Sn were used and data were taken
for all four combinations of projectile and target at 100 A MeV and for the two
projectiles and the $^{124}$Sn target at 150 A MeV.

\begin{figure}		

    \centering
    \epsfysize=8.0cm

   \epsffile{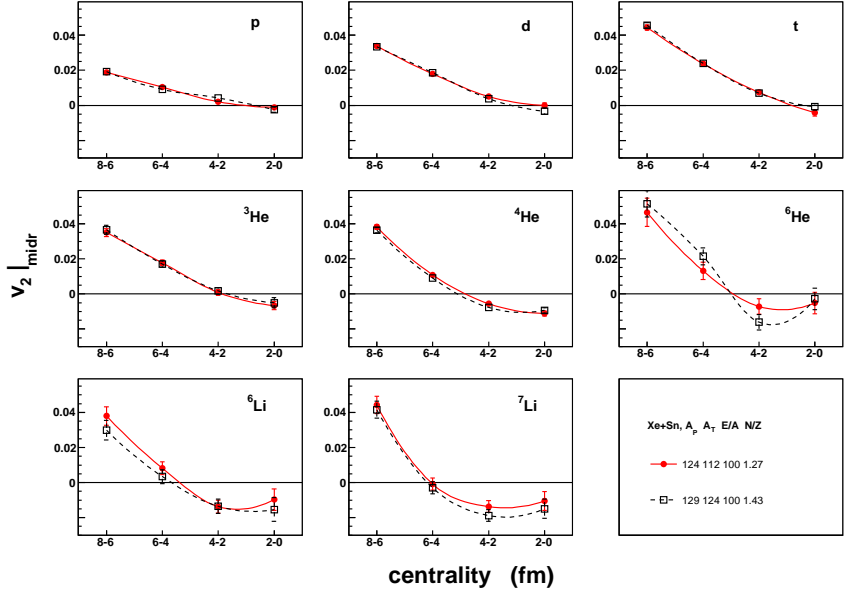}

\caption{Elliptic flow parameter $v_{2}$ at mid-rapidity for eight reaction
products with $Z \le 3$ integrated over $p_t/m > 0.1$ as a function of centrality
for the most neutron rich (open squares, black) and the most neutron poor
(dots, red) reaction systems at 100 A MeV.
}

\label{fig:v2}

\end{figure}

The INDRA multi-detector provides good acceptance and 
mass resolution at and near midrapidity at the present range
of bombarding energies. The flow parameters $v_1$ and $v_2$, appearing as 
coefficients of the Fourier decomposition

\begin{equation} \label{eq:def}
\frac{dN}{d(\phi-\phi_R)} = \frac{N_0}{2\pi}( 1+2 \sum_{n\ge 1} v_n {\rm cos} n(\phi-\phi_R))
\end{equation}
\noindent of the azimuthal yield distributions, were obtained from the corresponding
fits and determined as a function of impact 
parameter, rapidity $y$ and transverse momentum $p_t$. Examples of the obtained
results are shown for two, the most neutron-rich and the most neutron poor, 
of the four systems studied at 100 A MeV in Figs.~\ref{fig:v1} and~\ref{fig:v2}.
The dependence of the directed flow (slope $\partial v_1/ \partial y$ 
at mid-rapidity) and of
the elliptic flow ($v_2$ at mid-rapidity) are shown for eight reaction 
products with $Z \le 3$, selected with the condition $p_t/m > 0.1$.

The overall behaviour follows the expectations, in particular the dependence on the
fragment mass and the impact parameter (see figures) as well as that on the 
bombarding energy and on $p_t$ (not shown). A possibly isotopic effect is observed 
for the directed flow which is consistently larger for the more neutron-rich (but 
with 17 more nucleons also more massive) system. It is not obvious that a different
behaviour of mirror nuclei can be ascertained, 
and isotopic effects are practically absent
in the elliptic flow (Fig.~\ref{fig:v2}).

The displayed data are uncorrected which is sufficient 
to permit the direct visualization of the 
magnitude of the isotopic effects. For a comparison with theory, the above 
mentioned corrections will have to be applied or, alternatively, simulated events 
will have to be passed through a filter routine.

\end{document}